\documentclass[%
 aip,
 amsmath,amssymb,
preprint,%
groupedaddress,
]{revtex4-1}

\usepackage{graphicx}% Include figure files

\usepackage{dcolumn}% Align table columns on decimal point
\usepackage{bm}% bold math
\usepackage{ragged2e}
\usepackage[rightcaption]{sidecap}
\sidecaptionvpos{figure}{t}
\usepackage[utf8]{inputenc}
\usepackage[T1]{fontenc}
\usepackage{mathptmx}
\usepackage{etoolbox}
\usepackage[hidelinks]{hyperref}
\hypersetup{
  colorlinks   = true, %Colours links instead of ugly boxes
  urlcolor     = blue, %Colour for external hyperlinks
  linkcolor    = blue, %Colour of internal links
  citecolor   = blue %Colour of citations
}

\emergencystretch=\maxdimen
\usepackage{lmodern}
\usepackage{hyperref}

\newcommand*{\figref}[2][]{%
  \hyperref[{fig:#2}]{%
    Fig.~\ref*{fig:#2}%
    \ifx\\#1\\%
    \else
      #1%
    \fi
  }%
}

\makeatletter
\def\@email#1#2{%
 \endgroup
 \patchcmd{\titleblock@produce}
  {\frontmatter@RRAPformat}
  {\frontmatter@RRAPformat{\produce@RRAP{*#1\href{mailto:#2}{#2}}}\frontmatter@RRAPformat}
  {}{}
}%
\makeatother
\begin{document}

%\preprint{AIP/123-QED}

\title[Godsi et al.]{Exploring the Nature of High-Order Cavity Polaritons under the Coupling-Decoupling Transition}
% Force line breaks with \\
\author{M. Godsi}
 
\author{A. Golombek}%

\author{M. Balasubrahmaniyam}
%\homepage{http://www.Second.institution.edu/~Charlie.Author.}
%\affiliation{%
%Second institution and/or address%\\This line break forced% with \\
%}%

\author{T. Schwartz$^{\ast}$}
%\affiliation{ 
%Authors' institution and/or address%\\This line break forced with %\textbackslash\textbackslash
%}%
\affiliation{ 
School of Chemistry, Raymond and Beverly Sackler Faculty of Exact Sciences and Center for Light-Matter Interaction, Tel Aviv University, Tel Aviv 6997801, Israel.
%\\This line break forced with \textbackslash\textbackslash
}%
\email{talschwartz@tau.ac.il.}

\date{\today}% It is always \today, today,

\begin{abstract}
Recently, we predicted theoretically that in cavities which support several longitudinal modes, strong coupling can occur in very different manners, depending on the system parameters.
Distinct longitudinal cavity modes are either entangled with each other via the material or independently coupled to the exciton mode.
Here we experimentally demonstrate the transition between those two regimes as the cavity thickness is gradually increased while maintaining fixed coupling strength.
We study the properties of the system using reflection and emission spectroscopy and show that even though the coupling strength is constant, different behavior in the spectral response is observed along the coupling-decoupling transition.
In addition, we find that in such multimode cavities pronounced upper-polariton emission is observed, in contrast to the usual case of a single-mode cavity. Furthermore, we address the ultrafast dynamics of the multimode cavities by pump-probe spectroscopic measurements and observe that the transient spectra significantly change through the transition. 
\end{abstract}

\maketitle

\section*{INTRODUCTION}
When organic molecules are embedded inside an optical cavity, their interaction with light can be enhanced.
As the interaction becomes strong enough to overcome any decay processes, the system enters the strong coupling regime, where the wavefunctions of the molecules are modified and new quantum states, known as polaritons, are created~\cite{Ebbesen2016,Hertzog2019}. 
The coupling strength, which is described by the Rabi energy, is usually considered to be the only parameter that governs the physical behavior of the system, regardless of the dimensions of the cavity~\cite{Hertzog2020}.
In turn, the (collective) coupling strength generally depends on $\sqrt{N/V_c}$, where $V_c$ is the effective volume of the electromagnetic cavity mode and $N$ is the number of molecules within one mode~\cite{Ebbesen2016}.
Moreover, in many scenarios, the molecules are uniformly distributed within the cavity volume, in which case the Rabi energy simply scales as the square root of the molecular concentration.
Therefore, in such circumstances, it is often anticipated that the behavior of the strongly coupled cavity will solely rely on the molecular density.

However, in recent theoretical work~\cite{Balasubrahmaniyam2021}, we have predicted that multimode cavities, in which several longitudinal modes can reside close to the exciton transition energy~\cite{Coles2014,Richter2015, Simpkins2015, George2016, Li2020, Georgiou2021a}, can exhibit two fundamentally different coupling mechanisms depending on the system parameters.
In one, multiple cavity modes are mixed via their interaction with the material and form a mid-polariton branch in the cavity dispersion, while in the other mechanism, each one of the cavity modes couples independently to the exciton transition.
Here, each one of the cavity dispersion curves splits into a separate pair of (upper and lower) polariton branches, with the cavity modes remaining completely decoupled from each other, and with the system displaying a polaritonic band gap.
As we showed, below the transition the coupled system needs to be described by an $(N+1)$-dimensional Hamiltonian, whereas above the transition, a $2N$-dimensional Hamiltonian needs to be invoked in order to reproduce the decoupled polaritons dispersion.
Furthermore, our analysis has shown that the transition between these two types of strong coupling occurs at a critical length, which in turn depends on the molecular concentration.
Interestingly, at the critical length, the photon round-trip time, given by the cavity length divided by the group velocity in the medium, becomes comparable to the exciton lifetime~\cite{Balasubrahmaniyam2021}.
This theoretical observation indicates that the coupling-decoupling transition is related to the competition between the propagation of coherence across the system and its loss due to irreversible processes.
Hence, in contrast to the widespread assumption, multimode cavities having identical coupling strengths but different sizes are predicted to display different physical properties.

\begin{figure}
\includegraphics{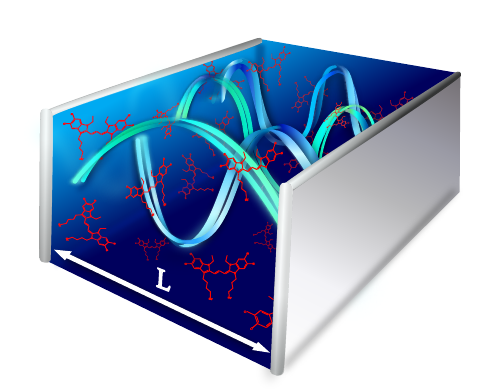}
\caption{
A sketch of the multimode cavity, consisting of two 35~nm Ag mirrors with a polymer spacer doped with TDBC molecules between them. The cavity length (thickness) $L$, is large enough such that the cavity contains several longitudinal modes. The first four modes are depicted.
}
\label{fig:Cavity_Sketch}
\end{figure}

The existence of these two different states within the strong coupling regime was confirmed experimentally, by varying either the coupling strength or the spatial distribution of the excitonic material within the cavity~\cite{Georgiou2021,Mandal2023a}.
Here, we directly explore this coupling-decoupling transition and demonstrate it experimentally by changing the cavity length, while maintaining a fixed coupling strength, showing that the behavior of the coupled system can be affected solely by its size.
We present a comprehensive spectroscopic study of polaritonic states in multimode cavities across the transition, employing both reflection and emission measurements.
Finally, we investigate the ultrafast dynamics of the system, using transient pump-probe spectroscopy.
\section*{RESULTS}
To examine the behavior of multimode strong coupling and its dependence on the system size, we fabricated a series of metallic cavities (\figref{Cavity_Sketch}) with increasing cavity thickness, in the range of $\sim400-1600$~nm, all doped with TDBC J-aggregates at a fixed concentration within a PVA host matrix
(for more details see Methods Section).
Each cavity was large enough such that it contained several longitudinal cavity modes in the vicinity of the molecular absorbance and the cavity thickness was adjusted so that at normal incidence the exciton energy resides exactly between two consecutive cavity modes (i.e., the cavity modes are equally detuned from the exciton but to opposite directions).
Using angle-resolved spectroscopy in reflection mode, we measured the dispersion of these multimode cavities, employing single-shot back focal-plane imaging, as described in the Methods Section.
The results (obtained for transverse electric polarization, as for all the measurements in this study) are presented in~\figref[(a)]{Reflection_Fig} for a cavity with $L = 628$~nm and~\figref[(b)]{Reflection_Fig} for $L = 1615$~nm.
In addition, the transfer matrix method (TMM) was used in order to fit these dispersion diagrams (shown as the black lines).
Repeating this procedure and averaging over the entire series of samples, we extracted a Rabi splitting energy of $\hbar\Omega_R = 128 \pm 5$~meV.
The critical thickness, namely, the length at which the system crosses the coupling-decoupling transition, was calculated under these conditions to be $ L_c\simeq 700$~nm (see Methods Section).
For the cavity with $L = 628$~nm, being shorter than the critical length (\figref[(a)]{Reflection_Fig}), the third- and forth-order cavity modes reside in the vicinity of the exciton energy $E_x = 2.1$~eV, with the third-order mode becoming resonant with the exciton at $k_x=7.7$~$ \mu$m$^{-1}$, where the avoided crossing and an energetic splitting is clearly observed (marked by the double-headed arrow).
Around $k_x = 0$, where $E_x$ lies exactly between the two modes, we clearly observe three dispersive features, which can be associated with a lower polariton branch (around 1.8~eV), an upper polariton branch (around 2.4~eV) and a mid-polariton branch around the exciton energy (2.1~eV).
As found in previous studies~\cite{Balasubrahmaniyam2021,Georgiou2021}, this mid-polariton mixes the exciton with the two nearby cavity modes, and in accordance we denote this branch as $| MP\,_{3-4} \rangle$.
In contrast, the lower-energy branch ($| LP\,_3 \rangle$) is formed by the hybridization of the third-order mode with the exciton and the higher-energy branch ($| UP\, _4 \rangle$) is formed by the hybridization of the forth-order mode with the exciton.
We note that the contributions of both cavity modes to the tripartite superposition composing $| MP\,_{3-4} \rangle$ are equal around $k_x = 0$.
When going to higher values of $k_x$, the photonic contribution to these polaritons will mostly come from the third-order cavity mode.
\begin{SCfigure}[][h!]
    \includegraphics{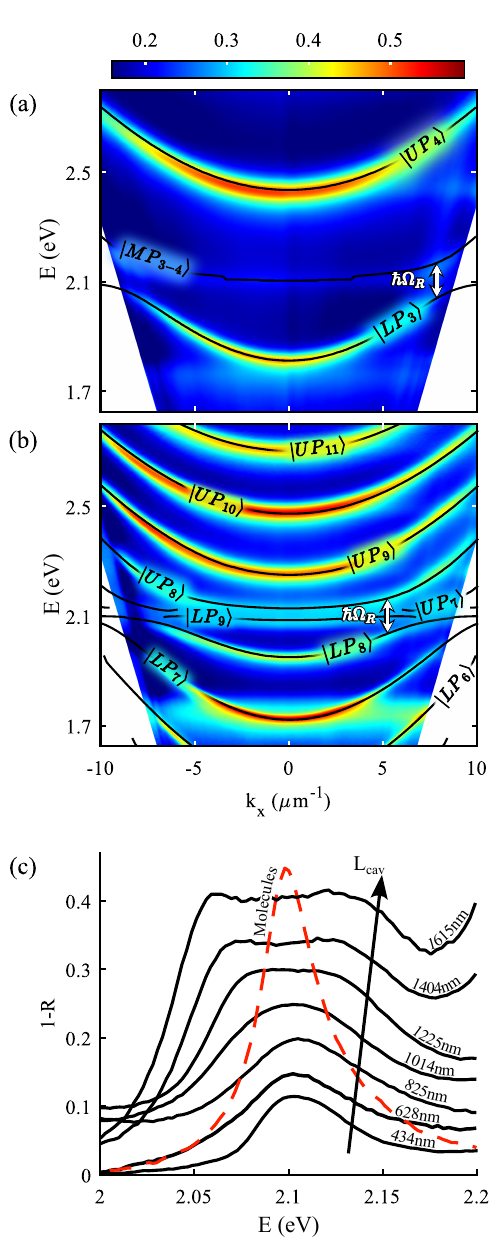}
    \caption{
    (a)-(c) Reflectivity measurements of the multimode cavities with varying lengths.
    (a) and (b) show the extracted dispersion maps of the cavities with $L=628$~nm and $L=1615$~nm, respectively (false-color map, corresponding to $1-R$ and measured for transverse electric polarization).
    The black solid lines show the results of the TMM simulation fitting and the location of the resonant anti-crossing is marked by the double-headed arrow, indicating the Rabi-splitting energy of $\sim 128$~meV.
    (c) $1-R$, measured at $k_x=0$ for multiple cavity lengths (black solid lines) around the bare exciton energy.
    The red dashed line corresponds to the bare film absorbance.
    }
    \label{fig:Reflection_Fig}
\end{SCfigure}
%
%\begin{figure}[h!]
    
%\end{figure}
%
The behavior of the system is very different for a cavity that is longer than the critical length, as appears in~\figref[(b)]{Reflection_Fig} for $L = 1615$ nm.
Here, the 8\textsuperscript{th} and 9\textsuperscript{th} longitudinal modes interact with the material, with their energies being equally detuned from $E_x$ at $k_x = 0$ (with opposite signs).
As a result, two independent sets of polaritonic branches, marked as $| LP\,_8 \rangle$, $| UP\,_8 \rangle\ $ and $| LP\,_9 \rangle$, $| UP\,_9 \rangle$ are created, with a band-gap appearing at $2.1$~eV between $| LP\,_9 \rangle$ and $| UP\,_8 \rangle\ $.
At $k_x = 5.4$~$ \mu$m$^{-1}$, where the dispersion of the 8\textsuperscript{th}-order mode crosses $E_x$, a clear anti-crossing is observed.
We stress that the observed splitting (marked by the double-headed arrow) is identical to the splitting observed in~\figref[(a)]{Reflection_Fig}, which is consistent with the fact that the molecular concentration is the same in both samples and that both cavities are completely and homogeneously filled with molecules.

By examining the normal-incidence ($k_x = 0$) reflection spectrum for a series of cavities with gradually increasing dimensions, we follow the evolution of the system as it passes through the coupling-decoupling transition, as shown in~\figref[(c)]{Reflection_Fig}.
We focus on the spectral region close to the bare TDBC absorption (shown as the red dashed line) and plot the inverted reflection spectrum ($1-R$), which is indicative of the absorption in the coupled cavity~\cite{Schwartz2013}.
When the cavity length is below the critical thickness we observe a mid-polariton branch (which appears as a single peak) that gradually broadens and splits into two peaks upon the creation of the decoupled polaritonic branches above the transition.
The qualitative changes that we observe in the spectral response of the cavities, despite having the same coupling strengths, clearly demonstrate how the system crosses the coupling-decoupling transition, highlighting the strong influence of the cavity dimensions on the physical behavior of the system.

Next, we explored the photoluminescence properties of the system.
Revealing the emission behavior of the multimode cavity provides us with a better understanding of various characteristics of the polaritons and has implications for energy transfer processes between the various states of the system.
As in the reflection measurements, we employed back-focal-plane imaging to measure the angle-resolved emission spectra, from which the dispersion was extracted (for details see the Methods Section).
We stress that each emission measurement was obtained at exactly the same spot as its respective reflection measurement shown in~\figref{Reflection_Fig}, to avoid minor differences due to sample inhomogeneity.
\figref[(a)]{Emission_Fig} shows the results for a cavity before the transition (same as in~\figref[(a)]{Reflection_Fig}, $ L= 628$~nm) and the dispersion shown in~\figref[(b)]{Emission_Fig} was measured for a cavity after the transition (corresponding to~\figref[(b)]{Reflection_Fig}, with $ L= 1615$~nm).
The TMM simulation (white lines) fits both the reflection and emission spectra.
In both cavities,~\figref[(a)]{Emission_Fig} and~\figref[(b)]{Emission_Fig}, the emission from the lower polariton branches is well observed, as reported in many previous studies~\cite{Lidzey1999,Schwartz2013,Coles2013,Wang2014}.
Surprisingly, the measurements showed emission from upper polariton branches as well, which is usually not observed in molecular systems due to their fast non-radiative decay.
\begin{SCfigure}[][h!]
\includegraphics{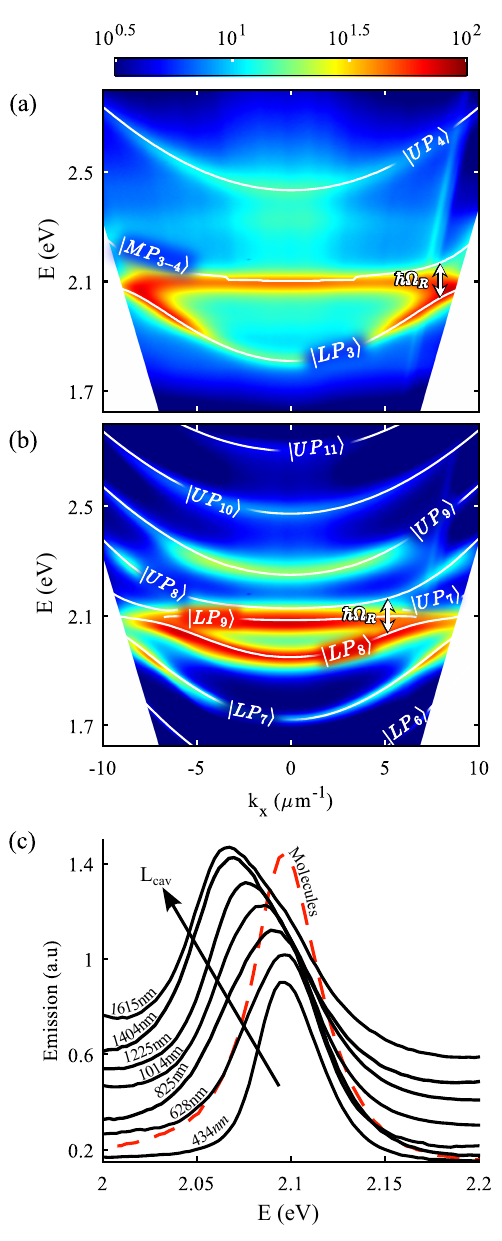}
\caption{
(a), (b) Dispersion maps, measured in emission mode, of the cavities with $L=628$ nm and $L=1615$ nm, respectively.
The white solid lines show the results of the TMM model fitting.
For comparison, the emission from a similar cavity without TDBC and under identical experimental conditions is shown in Appendix A.
(c) Emission at $k_x=0$, in the vicinity of the bare exciton energy, measured for cavities with increasing length (black solid lines).
The red dashed line shows the emission spectrum of the bare molecular film.
}
\label{fig:Emission_Fig}
\end{SCfigure}
In~\figref[(c)]{Emission_Fig} we explored the emission properties of the polaritons as the cavity length gradually varied.
In a similar manner to~\figref[(c)]{Reflection_Fig}, we plot in~\figref[(c)]{Emission_Fig} the emission spectra in the vicinity of the exciton energy at $k_x=0$.
When the cavity length was below the critical length, we clearly observed emission from the mid-polariton branch (two lowest curves, corresponding to $|MP\,_{2-3}\rangle$ and $|MP\,_{3-4}\rangle$ for the $434$~nm and $628$~nm cavities, respectively).
Interestingly, as the length of the cavity was increased and the mid-polariton branch transitioned to a pair of separated polaritonic branches ($| UP \rangle$ and $| LP \rangle$), the emission was observed only from the lower state of each pair. 
This behavior becomes more pronounced for thicker cavities, for which the polaritonic band gap widens, with the $| LP \rangle$ emission peak gradually shifting to lower energies.
The missing emission from the higher-energy state may be the result of the fast relaxation from the upper polaritons to the lower polaritons (which is reminiscent of the behavior in normal single-mode cavities).
However, here the energy separation between the two polaritons is rather small and comparable to $k_B T$, which indicates that such a direct relaxation mechanism should not result in such a strong preference for emission from the lower-energy state.
For example, in~\figref[(b)]{Reflection_Fig}, the energy difference between $| LP\,_9 \rangle$ and $| UP\,_8 \rangle$ is about 40~meV.
A possible explanation for such preference may arise from the different spatial symmetries of the polaritonic states.
For example, $| LP\,_8 \rangle$ and $| UP\,_8 \rangle$ both have the same spatial distribution within the cavity volume, namely, 8 antinodes between the mirrors.
Conversely, $| LP\,_9 \rangle$ and $| UP\,_9 \rangle$ both have 9 antinodes~\cite{Balasubrahmaniyam2021}.
These symmetry considerations may explain why energy relaxation from $| UP\,_9 \rangle$ to $| UP\,_8 \rangle$ is blocked, leading to preferential relaxation to $| LP\,_9 \rangle$.

\begin{figure}[h!]
    \includegraphics{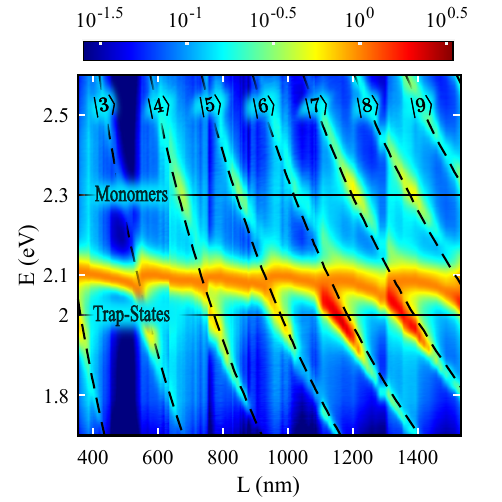}
    \caption{Emission (log scale and normalized with respect to the MP emission signal) cross-sections at $k_x=0$ as a function of the cavity length, measured at intervals of 10 nm.
    The horizontal black lines represent the self-trapped excitons of the TDBC J-aggregates at $E=2$~eV and the monomer emission at $E=2.3$~eV.
    The dashed black lines show the TMM model fits empty cavity modes.}
    \label{fig:thicknesses}
\end{figure}

In order to further explore the emission properties of the system and their dependence on the cavity detuning, we now plot in~\figref{thicknesses} the emission intensity (measured at $k_x=0$) as a function of energy and cavity thickness (which is now varied by small steps of about 10 nm).
By rearranging the data in this way, we are able to look at the emission properties of all the high-order cavity polaritons at once, which allows us to observe some very interesting behavior.
As shown in~\figref{thicknesses}, the emission from the cavities appears to be enhanced in the vicinity of specific locations, exhibiting some periodic trend with the cavity thickness.
The enhanced emission occurs when the upper polariton branches overlap with the TDBC monomer at $E=2.3$~eV, and also whenever the lower polariton branches cross an energy value of $\sim 2.0$~eV, which was shown to correspond to self-trapped states in the J-aggregate~\cite{Sorokin2019}.
Interestingly, this effect becomes much more pronounced for thicker cavities ($L>1000$~nm), i.e., above the coupling-decoupling transition and after the mid-polariton splits into two polaritonic branches.
We stress that clear emission signals are observed also away from these locations, both from the lower polariton branches as well as the upper polaritons.

\begin{figure*}
    \centering
    \includegraphics[width=\textwidth]{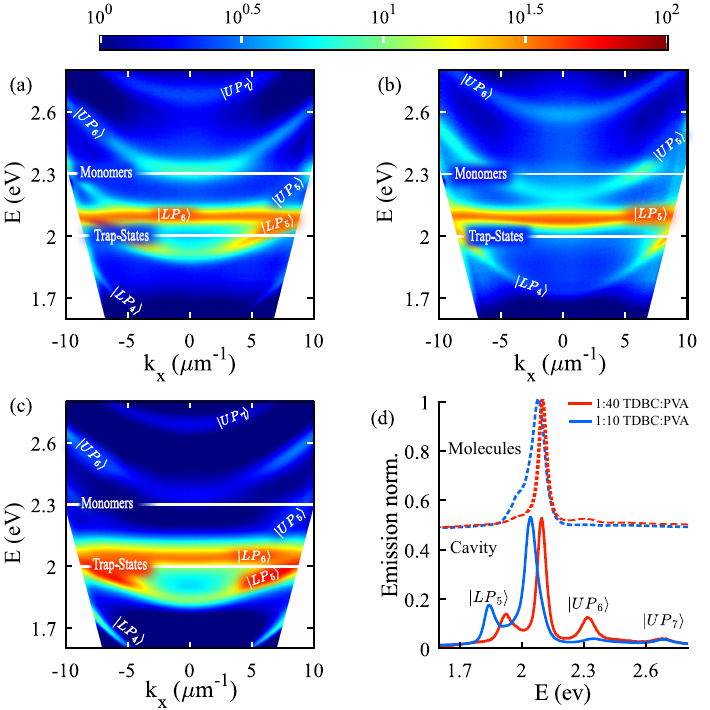}
    \caption{(a)-(c) Emission dispersion spectra (log scale) normalized to the mid-polariton. (a) and (b) have the same molecular concentration, TDBC:PVA 1:40, and cavity lengths of $L=1022$~nm and $L=930$~nm, respectively. (a) and (c) have the same cavity length $L=1022$~nm but (c) has a TDBC:PVA ratio of 1:10. The solid white lines represent the energy of the self-trapped excitons at $E=2$~eV and the monomers at $E=2.3$~eV. (d) Emission spectra of molecules inside (solid lines) and outside (dashed lines) of the cavity. The dashed lines are shifted upward for clarity. The red lines represent a TDBC:PVA ratio of 1:40 and the blue lines represent a TDBC:PVA ratio of 1:10.}
    \label{fig:four}
\end{figure*}

To further examine the emission properties of the multimode cavity and the interplay between the polariton emission and the molecular spectral features, we compare the full dispersion maps for two slightly different thicknesses, as shown in~\figref[(a)]{four} and~\figref[(b)]{four}, with $L=1022$~nm and $L=930$~nm, respectively (both above the coupling-decoupling transition).
As discussed above, we again observe the increase of the lower polariton emission ($| LP\,_5 \rangle\ $ in~\figref[(a)]{four} and $| LP\,_4 \rangle\ $ in~\figref[(b)]{four}) emission intensity once its dispersion curve crosses the energy of the self-trapped excitons at $E=2$ eV (marked by the lower white solid line).
Furthermore, we observe a maximum in the emission intensity when the upper polariton dispersion coincides with the monomer emission at $E=2.3$~eV ($| UP\,_6 \rangle\ $ in \figref[(a)]{four} and $| UP\,_5 \rangle\ $ in \figref[(b)]{four}).
As expected, the various crossing points occur at different values of the in-plane momentum $k_x$ for the two different cavities.
For example, in~\figref[(a)]{four} the $| UP\,_6 \rangle\ $ intersects with the monomer emission line at $k_x=0$, while in~\figref[(b)]{four} $| UP\,_5 \rangle\ $ intersects with the monomers at $|k_x|>0$.
However, the magnitude of the observed enhancements, both around $2.3$~eV and $2.0$~eV, is identical for the two samples.
Overall, our measurements demonstrate that the emission properties of the coupled system are strongly affected by all the molecular states, even when they are not strongly coupled with the cavity.
In particular, these results suggest that the observed enhancement originates from efficient energy transfer between the molecular states and the polaritons when they are resonant with each other.

To isolate the contribution of the monomers and to reveal the pure emission profile of the upper polariton, we fabricated a cavity that has a higher TDBC concentration (with a 1:10 TDBC/PVA mass ratio) and with the same cavity length as the cavity in~\figref[(a)]{four} (L=1022 nm).  
As we increase the concentration of the molecules, the formation of J-aggregates is promoted, reducing the concentration of residual monomers~\cite{Wurthner2011}.
When measuring the emission from this sample (see~\figref[(c)]{four}), we observe that the enhancement around the monomer energy ($2.3$~eV) no longer occurs, while clear dispersive features, matching the upper polaritonic branches $| UP\,_6 \rangle\ $and $| UP\,_7 \rangle\ $, are still visible.
Note that the critical length for the coupling-decoupling transition is reduced when the molecular concentration is increased.
In a similar manner to the behavior seen in~\figref{thicknesses}, we again find strong emission enhancement in the vicinity of $2.0$~eV (related to energy transfer from the trap-states, and here occurring around $k_x \sim 6 \mu$m$^{-1}$), but now at a lower cavity thickness.
However, due to the higher concentration and lower critical length, this sample resides even deeper in the decoupled regime (compared with the samples shown in~\figref[(a)]{four}, which has the same thickness).
This behavior provides a further indication of the connection between the enhancement effect and the coupling-decoupling transition.

\begin{SCfigure}[][h!]
\includegraphics{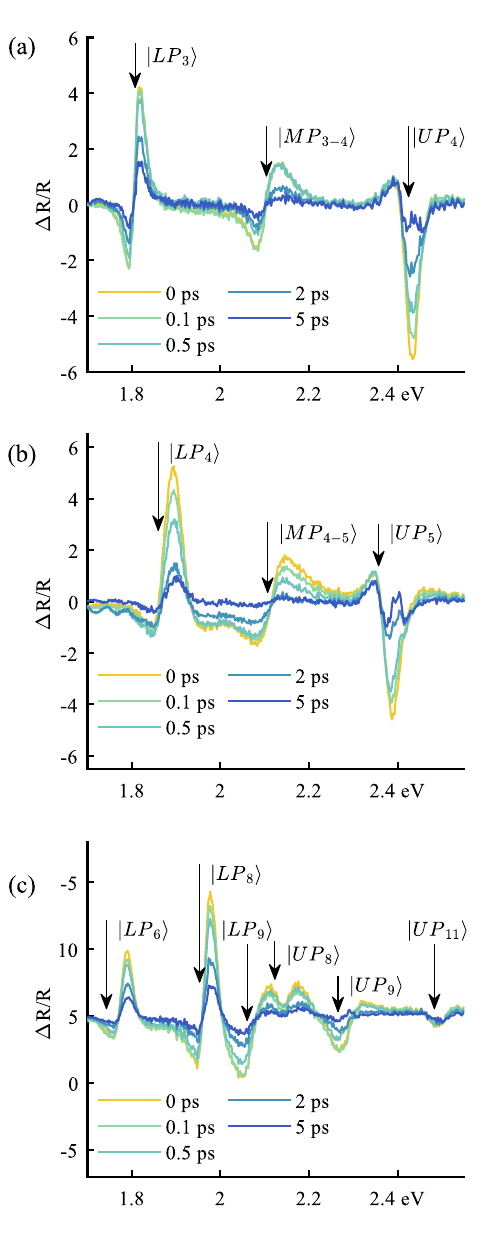}
\caption{Transient reflectance spectra immediately after pumping at the respective energies of the upper polaritons, (a) $| UP\,_4 \rangle\ $ (b) $| UP\,_5 \rangle\ $ and (c) $| UP\,_9 \rangle\ $, and their evolution as a function of time. The cavity lengths are (a) $L=628$ nm (b) $L=825$ nm and (c) $L=1615$ nm.}
\label{fig:transient_spectra} 
\end{SCfigure}

To allow better quantitative evaluation of these effects, we compare in~\figref[(d)]{four} the (zero-angle) cavity emission (solid lines) with the emission spectra of the bare molecular films at both concentrations (dashed lines).
It is worthwhile noting that for the bare film emission, the disappearance of the monomer emission peak when the TDBC concentration is increased, while, at the same time, the contribution of the self-trapped states becomes stronger.
We also note that, in the cavities and at both concentrations, we observe comparable emission intensity from the $| UP\,_7 \rangle\ $, irrespective of the monomer concentration, which is also comparable to the $| UP\,_6 \rangle\ $ peak in the 1:10 sample.
In a previous study, similar features were attributed to emission from TDBC monomers~\cite{Coles2014}.
However, the data presented in~\figref{four}, which compares the emission spectra and their dispersion from the two multimode cavities with different concentrations, confirms that the signals we observe must be associated with upper polariton emission (which can be influenced to some extent by the monomers, as discussed above).
Interestingly, this upper polariton emission is much higher than what is observed in normal single-mode (thin) cavities~\cite{Coles2011, Coles2011a, Schwartz2013}.

Next, we investigated the ultrafast dynamics of the multimode cavity along the coupling-decoupling transition, using transient pump-probe spectroscopy (in reflection mode, see Methods Section).
\figref{transient_spectra} displays the temporal evolution of the differential reflectivity spectra $\Delta{R}/R$ following the excitation of the upper polariton state for three cavities having different thicknesses ($628$, $825$, and $1615$~nm) residing below, near, and above the coupling-decoupling transition.
Generally, the lower polariton signals have an S shape (negative dip and positive peak), similar to previous transient absorption studies of polaritons~\cite{Virgili2011, Schwartz2013}, and in turn, the upper polaritons display a negative dip.
These transient signals maintain their shapes along the transition, differing primarily in their spectral positions, as a result of the different mode-orders which are coupled with the molecules.

\begin{SCfigure}[][h!]
\includegraphics{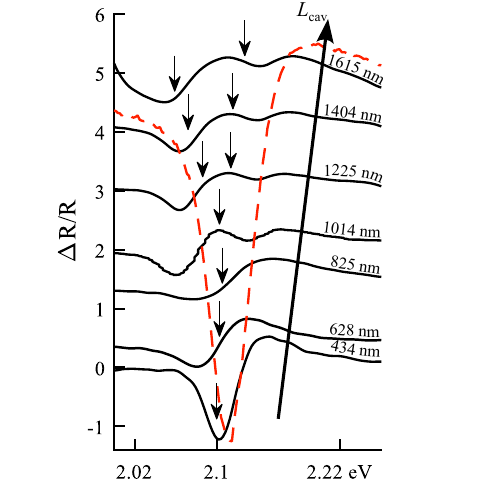}
\caption{Transient reflectance spectra immediately after pumping at the respective energies of the upper polaritons, in the region of the TDBC transient spectrum (red dashed line). From the bottom graph to the top graph the cavity
length increases from $434$ to $1615$ nm. The arrows on the graphs indicate the position of the mid-polariton/polaritonic pair in the reflection (static) spectrum (\figref[(c)]{Reflection_Fig}).}
\label{fig:Transient_zoom}
\end{SCfigure}

In contrast, the spectral shape of the middle branch does change throughout the coupling-decoupling transition.
Below the transition, as shown in~\figref[(a)]{transient_spectra}, the mid-polariton $| MP\,_{3-4} \rangle$ displays an S shape, similar to the LP's spectral shape.
Close to the critical length (see~\figref[(b)]{transient_spectra}) we observe a broadening of the transient signal $| MP\,_{4-5} \rangle$ and as the thickness is further increased, and a band gap forms between $| LP\,_{9} \rangle$, $| UP\,_{8} \rangle$, the signal in the vicinity of the molecular energy changes significantly (see~\figref[(c)]{transient_spectra}).  
Instead of an S shape, as seen for the mid-polariton signal, we observe a negative dip and two positive peaks.

\begin{SCfigure}[][h]
\includegraphics{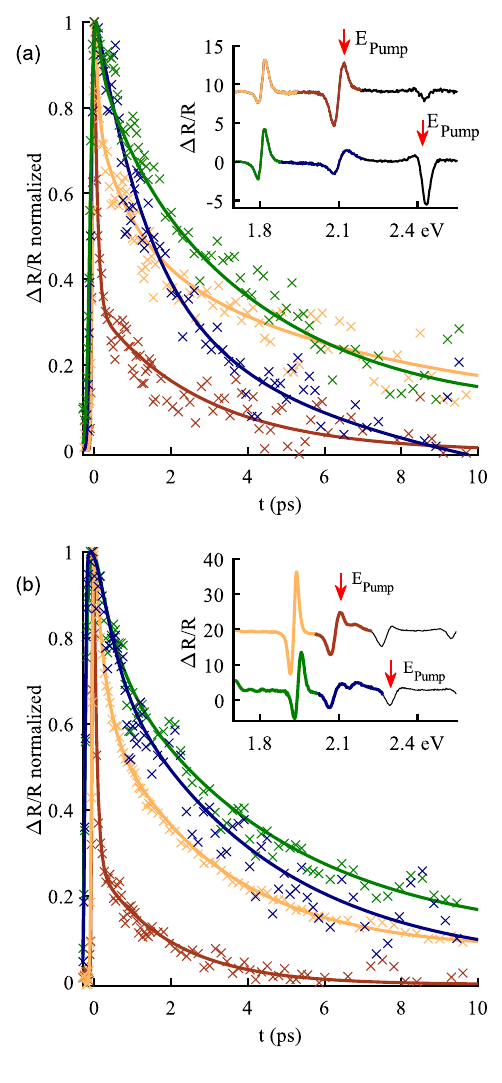}
\caption{
Normalized pump-probe kinetics measured (a) for the $L=628$~nm cavity (before the transition, with a mid-polariton branch at $2.1$ eV) and (b) for the $L=1404$~nm cavity (above the transition, where the mid-polariton splits to form a polaritonic bandgap).
The crosses represent the measured data, while the solid lines indicate a biexponential fit used as a guide to the eye.
Color coding is provided by the inset, which also shows the line-shapes of the various spectral regions in the transient signal, and the pump energies, marked by the red arrows.
Yellow and green correspond to the kinetics around the lower polariton energy and red and blue correspond to the kinetics around $2.1$ eV (mid polariton in (a) or polaritonic bandgap in (b)).
The pump energy was tuned to either the upper polariton (green and blue) or the mid-polariton/polaritonic bandgap (yellow and red).
}
\label{fig:dynamic}
\end{SCfigure}

In~\figref{Transient_zoom} we present the differential reflectivity spectra immediately after the excitation pulse, measured for the entire series of samples as in~\figref[(c)]{Reflection_Fig} and~\figref[(c)]{Emission_Fig}, focusing on the spectral region around the mid-polariton.
As a reference, we also plot the transient reflection spectrum measured for a bare TDBC/PVA film (deposited on a mirror).
The arrows on the graphs mark the location of the mid-polariton/polaritonic pair as obtained from the static measurement (\figref[(c)]{Reflection_Fig}).
The substantial change in the spectral shape when the cavity length increases is well observed, which emphasizes once again the strong effect of the cavity dimensions on the cavity properties in general and particularly on the dynamic profile. 

In~\figref{dynamic} we explore the decay kinetics of the system before (\figref[(a)]{dynamic}, $L=628$~nm) and after (\figref[(b)]{dynamic}, $L=1404$~nm) the transition.
For each case, we compare the kinetics at spectral regions around the lower polariton branch with those around the mid-polariton/band-gap pair.
This comparison is made following an excitation pulse with energy corresponding to either the upper polariton energy ($2.4$~eV in (a) and $2.3$~eV in (b)) or the bare-molecules energy, $2.1$~eV, (see inset for color coding).
In most cases, the signals exhibit typical decay times of $1.5-3$ ps, which is slower than the decay of bare molecules (with $\tau_{1/2}\sim0.5$~ps), as previously seen in single-mode TDBC cavities~\cite{Virgili2011, Schwartz2013,Mewes2020}.
Interestingly, however, both the mid-polariton (red curve in~\figref[(a)]{dynamic}) and the band-gap pair (red curve in~\figref[(b)]{dynamic}) display a much faster decay when the pump energy is at $2.1$~eV, with a half-life time of $\sim 100$ fs.
While a more systematic exploration is required to understand the observed dynamics, these preliminary results indicate that in multimode cavities, the existence of polaritonic states in the vicinity of the bare-molecule energy strongly influences the decay kinetics, possibly by modifying the energy exchange with the dark-states manifold.

\section*{CONCLUSIONS}
In conclusion, we presented the experimental study of strong coupling in a series of multimode cavities and successfully demonstrated the coupling-decoupling transition with the cavity length.
Specifically, we showed that the dimensions of the cavity, which are usually ignored in cavity quantum electrodynamics, influence the behavior of strongly coupled systems, even when the light-matter coupling strength remains fixed.
We comprehensively explored the emission properties of this system throughout the coupling-decoupling transition and found that below the transition the mid-polariton emits while above the transition, where the mid-polariton splits into two decoupled states, only the lower of them is observed to be emissive.
We distinctly observed upper polariton emission in those multimode, strongly coupled cavities, which is found to be much stronger than in the usual case of single-mode cavity.
At the same time, we revealed that the presence of monomers results in a significant enhancement in the upper polariton emission when their energies overlap, suggesting energy transfer from the monomers to the UP states.
Moreover, we found that the presence of self-trapped states in the molecular J-aggregates affects the lower polariton emission in a similar manner.
Finally, we investigated the ultrafast dynamics of the system and showed clear evidence that the transient spectra are modified through the transition.
However, further investigation is required in order to obtain a better understanding of such systems and to fully resolve the energy relaxation mechanism in such strongly coupled multimode cavities.
Our observations open up a new opportunity for engineering the photophysical properties of molecules, specifically in systems that employ strong coupling for long-range intermolecular energy transfer.

\section*{METHODS}
We fabricated a series of cavities with increasing cavity thickness while maintaining a fixed molecular concentration. 
Following a similar procedure as in previous studies~\cite{Golombek2020}, each one of the cavities was based on a 35 nm thick silver layer that was sputtered onto a glass substrate.
On top of that, a PVA film doped with TDBC dye molecules (5,6-Dichloro-2-[[5,6-dichloro-1-ethyl-3-(4-sulfobutyl)-benzimidazol-2-ylidene]-propenyl]-1- ethyl-3-(4-sulfobutyl)-benzimidazolium hydroxide, inner salt,
sodium salt, Few Chemicals) was deposited by spin coating, followed by the deposition of a second 35 nm thick silver film as the top mirror.
The PVA/TDBC solution was prepared by dissolving PVA ($205,000$ MW, Aldrich) in water (with 5\% w/w concentration) at $110^\circ$C, followed by adding the TDBC powder directly into the PVA solution, with a 1:40 weight ratio between TDBC and PVA.
This final solution was stirred at $70^\circ$C for one hour.
The spin coating speed was varied in the range of $300-1500$ rpm, providing a layer thickness range of $\sim 400-1600$~nm.

We measured the reflection and emission dispersion spectra of the multimode cavities using an inverted microscope (Olympus IX71, 60X objective) paired with an imaging spectrometer (IsoPlane SCT 320, Princeton Instruments).
Both types of measurements were performed at identical locations on the samples, over an area of $\sim 110$ $\mu$m.
For the reflection measurements, we used a halogen lamp, and for the emission measurements, the excitation was done using a mercury lamp with a band-pass filter of $330-385$~nm, and the emission was filtered with a 420 nm long-pass filter.
In both cases, the polaritonic dispersion was obtained by back focal plane imaging using an 8 cm lens placed between the microscope output port and the spectrometer. Additionally, a polarizer was placed after the lens, such that polaritons with TE polarization were measured.

For the measurements of the ultrafast dynamic, we used an fs pump-probe setup (Helios, Ultrafast Systems), operated in reflection mode at normal incidence.
For the pump, we used an 80 fs, 1 KHz, tunable-wavelength OPA (Topas, Light Conversion), and the probe was a white-light pulse ($400-800$ nm) generated by a sapphire plate.
All the measurements were chirp-corrected using a standard software procedure.

To calculate the theoretical value of the critical cavity length $L_c$, we numerically solved the implicit equation $L_{c}=\frac{hc(n_{0}\gamma)}{\pi\left[1-2\beta\right]\left[f-(n_{0}\gamma)^{2}\right]}$ with $\beta = \left(e^{\frac{2\pi f L_c}{h c n_0 \gamma}} -1 \right)^{-1}$, $n_0 = 1.5$ being the background refractive index inside the cavity, $\gamma = 34$~meV$^2$ the FWHM of the molecular absorption spectrum, $c$ the speed of light in vacuum, $h$ the Planck constant and $f$ the oscillator strength per unit volume.
Following the same definition as in Ref.~\citenum{Balasubrahmaniyam2021}, the oscillator strength can be related to the absorption spectrum by its Lorentz-type approximation $n(E)^2 = n_0^2 + f/\left(E_x^2 - E^2 -i \gamma E \right)$.
From fitting the TMM simulations to the measured dispersion curves (e.g.~\figref[(a)]{Reflection_Fig}) we obtained a value of $f \simeq 37$~meV$^2$, which gave a critical cavity length of $L_c \simeq 700$~nm.

\begin{acknowledgments}
This research was supported by the Israel Science Foundation, grant number 1435/19.
\end{acknowledgments}

%\section*{Data Availability Statement}
%
%AIP Publishing believes that all datasets underlying the conclusions of the paper should be available to readers. Authors are encouraged to deposit their datasets in publicly available repositories or present them in the main manuscript. All research articles must include a data availability statement stating where the data can be found. In this section, authors should add the respective statement from the chart below based on the availability of data in their paper.

\appendix

\section{Empty Cavity Emission}
\begin{figure}[h!]
\includegraphics{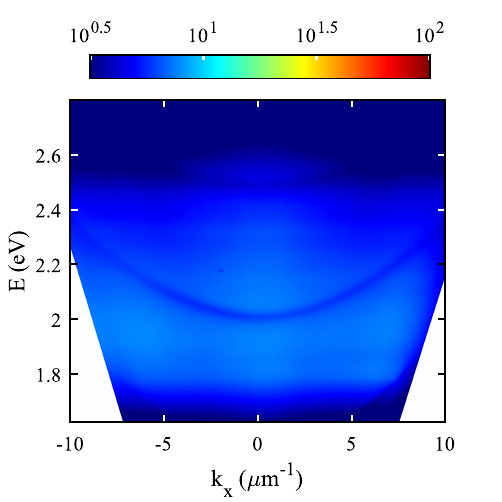}
\caption{Dispersion maps, measured in emission
mode, of an empty cavity (cavity without molecules) with $L=770$~nm.
The scale is the same as in~\figref{Emission_Fig}.}
\label{fig:Empty}
\end{figure}
To verify that the appearance of upper-polariton emission in ~\figref{Emission_Fig} does not originate from the leaking of background emission through the coupled cavity modes, we measured the emission from a cavity with a similar thickness under identical excitation intensity and identical integration time. As seen in~\figref{Empty}, the residual emission signal is negligible in comparison to the detected upper polariton emission.
%

%\nocite{*}
%\section*{REFERENCES}
\bibliography{ZoteroReferences}% Produces the bibliography via BibTeX.

\end{document}